\documentclass[journal,twoside,web]{ieeecolor}
\usepackage{generic}
\usepackage{array}
\usepackage[caption=false,font=normalsize,labelfont=sf,textfont=sf]{subfig}
\usepackage{textcomp}
\usepackage{stfloats}
\usepackage{url}
\usepackage{verbatim}
\usepackage{graphicx}
\usepackage{cite}
\usepackage[backref]{hyperref}
\usepackage{amsmath,amssymb,amsfonts,bm}
\usepackage{dsfont}              

\usepackage[ruled,vlined]{algorithm2e} 

\usepackage{booktabs}
\usepackage{threeparttable}
\usepackage{multirow}
\graphicspath{{./figures/}}
\newtheorem{theorem}{Theorem}

\def\BibTeX{{\rm B\kern-.05em{\sc i\kern-.025em b}\kern-.08em
    T\kern-.1667em\lower.7ex\hbox{E}\kern-.125emX}}
\begin{document}
\title{Probabilistic Forecasting Method for Offshore Wind Farm Cluster under Typhoon Conditions: a Score-Based Conditional Diffusion Model}
\author{Jinhua He, \IEEEmembership{Graduate Student Member, IEEE}, Zechun Hu, \IEEEmembership{Senior Member, IEEE}
\thanks{The authors are with Department of Electrical Engineering, Tsinghua University, Beijing 100084, China. (e-mail: zechhu@tsinghua.edu.cn).}
}

\maketitle

\begin{abstract}
Offshore wind power (OWP) exhibits significant fluctuations under typhoon conditions, posing substantial challenges to the secure operation of power systems. Accurate forecasting of OWP is therefore essential. However, the inherent scarcity of historical typhoon data and stochasticity of OWP render traditional point forecasting methods particularly difficult and inadequate. To address this challenge and provide grid operators with the comprehensive information necessary for decision-making, this study proposes a score-based conditional diffusion model (SCDM) for probabilistic forecasting of OWP during typhoon events. First, a knowledge graph algorithm is employed to embed historical typhoon paths as vectors. Then, a deterministic network is constructed to predict the wind power under typhoon conditions based on these vector embeddings. Finally, to better characterize prediction errors, a denoising network is developed. At the core of this approach is a mean-reverting stochastic differential equation (SDE), which transforms complex error distributions into a standard Gaussian, enabling the sampling of forecasting errors using a reverse-time SDE. The probabilistic forecasting results are reconstructed by combining deterministic forecasts with sampled errors. The proposed method is evaluated using real-world data from a cluster of 9 offshore wind farms. Results demonstrate that under typhoon conditions, our approach outperforms baseline models for both deterministic and probabilistic metrics, verifying the effectiveness of the approach.
\end{abstract}

\begin{IEEEkeywords}
Probabilistic forecasting, offshore wind farm cluster, typhoon conditions, score-based diffusion model, knowledge graph
\end{IEEEkeywords}

\section{Introduction}
\label{sec:introduction}
\IEEEPARstart{W}{ith} the advancement of technology and the reduction in costs, the installed capacity of offshore wind power has grown rapidly in recent years. The annual average utilization hours of offshore wind power are generally higher than those of onshore wind power \cite{heier2014grid}. However, the inherent stochasticity of wind power, exacerbated by extreme meteorological events such as typhoons, presents formidable challenges for accurate wind power forecasting (WPF) \cite{liu2024bayesian}. Existing studies primarily focus on WPF under normal weather conditions, with limited research addressing the unique complexities introduced by typhoons. During typhoon events, offshore wind turbines experience drastic power fluctuations—initial surges to rated capacity followed by abrupt shutdowns triggered by safety protocols due to excessive wind speeds \cite{10858410}. These dynamics, coupled with the scarcity of historical typhoon-specific data, render conventional forecasting models inadequate for capturing the non-stationary spatiotemporal correlations \cite{liu2023spatio}.

The chaotic nature of atmospheric systems under typhoon conditions necessitates probabilistic frameworks to quantify forecasting uncertainty \cite{cui2017data,arora2022probabilistic}. Existing approaches to modeling joint probability distributions can be broadly categorized into decoupled and mapping-based paradigms. Decoupled modeling decomposes the joint probability distribution into marginal distributions and correlation structures, typically assuming parametric forms such as Gaussian mixtures \cite{cui2017data} or Copula functions \cite{wang2019forecasted}. However, these methods struggle to generalize under typhoon conditions due to the non-Gaussian nature of prediction errors and the high-dimensional spatiotemporal dependencies among wind farms \cite{tang2024spatial,liu2023spatio}. Alternatively, mapping-based approaches leverage generative models to approximate the joint probability distribution without explicit parametric assumptions. Generative adversarial networks (GANs) \cite{ye2024novel} circumvent likelihood computation through adversarial training, but suffer from instability. Normalizing flows (NF) \cite{wen2022continuous} optimize the likelihood function via the Jacobian matrix, yet impose constraints on network structure and dimensionality. Variational autoencoders (VAEs) \cite{bond2021deep} employ the evidence lower bound (ELBO) for likelihood approximation but are limited by the expressiveness of the variational posterior distribution. Recently, diffusion models \cite{ho2020denoising}, particularly score-based diffusion models \cite{song2020score}, have emerged as promising alternatives, offering GAN-level sample quality without adversarial training, flexible architectures, exact log-likelihood computation, and the capability to solve inverse problems without retraining. However, their application to probabilistic WPF under typhoon conditions remains underexplored, particularly in addressing the following two key challenges:

1) Data Scarcity: Typhoon events are rare and geographically localized, resulting in limited training data. Neural networks typically require enough data to generalize effectively, making robust model training challenging under data constraints \cite{zhang2021understanding}.

2) Conditional Probability Modeling: While score-based diffusion models excel in generating high-quality unconditional samples, their capability to model probability distributions under specific conditions \cite{yi2024towards} remains underdeveloped—an essential requirement for probabilistic WPF under typhoon conditions.

To address the first challenge, techniques such as meta-learning \cite{meng2024adaptive} and transfer learning \cite{liu2021short} offer potential solutions but introduce additional training complexity. Recent advancements in few-shot learning emphasize domain-specific knowledge integration to enhance generalization. For instance, Bayesian deep learning has been successfully applied to model blade icing dynamics in cold climates, demonstrating improved robustness under data-limited conditions \cite{liu2024bayesian}. Similarly, reference \cite{yu2022ultra} proposed an ultra-short-term wind power subsection forecasting method that identifies sharp rise and fall patterns under extreme weather. However, typhoon-specific applications remain sparse, with existing studies often overlooking the temporal evolution of typhoon paths, leaving a critical research gap in probabilistic WPF.

For the second challenge, recent advances in controllable generation offer possibilities for WPF under typhoon conditions. In \cite{dong2023short}, a novel conditional latent diffusion model leveraging meteorological covariates for wind power scenario generation was proposed. Similarly, reference \cite{wen2022continuous} introduced a probabilistic forecasting approach using conditional normalizing flows, where numerical weather prediction (NWP) wind speeds serve as conditioning inputs. Reference \cite{dong2022data} designed a data-driven renewable scenario generation framework by embedding interpretable characteristic features in the latent space. These methods typically incorporate conditioning variables alongside auxiliary features into neural networks to directly generate probabilistic forecasts under specified conditions. However, for probabilistic WPF under typhoon conditions, direct modeling of the relationship between input conditions and output distributions remains highly challenging due to data scarcity.

To overcome these challenges, this paper proposes a score-based conditional diffusion model (SCDM) tailored for probabilistic WPF under typhoon conditions. To the best of the authors' knowledge, this is the first probabilistic forecasting modeling framework for offshore wind power cluster in real power systems that accounts for typhoon events. The key contributions of this work are summarized as follows:
\begin{enumerate}
\item Typhoon path embedding: Historical typhoon trajectories encapsulate valuable spatiotemporal signatures. In this work, 11 years of historical typhoon paths are encoded into dynamic feature vectors using a knowledge graph algorithm, enabling the extraction of typhoon-phase-specific information. As a form of domain-specific knowledge, this embedding facilitates probabilistic WPF under typhoon conditions, enhancing predictive accuracy while avoiding the additional training burden associated with meta-learning or transfer learning.
\item Deterministic-probabilistic hybrid architecture: To reduce the complexity of the denoising process and enhance sample quality with fewer diffusion steps, the forecasting task is decomposed into two components: deterministic forecast regression and forecast error sample generation. These components are handled by a deterministic network and a denoising network, respectively. The deterministic forecast, combined with the typhoon path embedding, serves as the input to the denoising network. 
\item Score-based conditional diffusion with mean-reverting SDE: To accurately model prediction errors while enabling controllable generation, this work introduces a score-based conditional diffusion model centered on mean-reverting SDEs. The proposed model progressively maps the complex error distribution into a standard Gaussian through a forward diffusion process, and subsequently samples prediction errors via a reverse-time SDE, ensuring robust uncertainty quantification under typhoon conditions.
\end{enumerate}

The remainder of this paper is organized as follows. Section \ref{sec:score-based} provides background on score-based diffusion models. Section \ref{sec:conditional} details theories of score-based conditional diffusion models and the corresponding training and sampling algorithms. Section \ref{sec:case_study} discusses numerical results. Finally, Section \ref{sec:conclusion} concludes the paper.

\section{Score-based Diffusion Models}
\label{sec:score-based}
\subsection{Problem Formulation}
Probabilistic WPF under typhoon conditions presents significant challenges. From the data perspective, in addition to relying on NWP, which are essential under normal conditions, additional auxiliary forecasts—such as typhoon path predictions—are required. From a modeling standpoint, a one-step generation approach often results in low-quality samples that fail to accurately reflect real-world situations. To address this, we adopt a two-stage strategy: first, a deterministic prediction is performed, followed by the generation of prediction errors. The probabilistic forecasts are then obtained by superimposing the sampled errors onto the deterministic prediction results. Mathematically, this can be depicted in formula \eqref{eq:formulation}.
\begin{equation}
    {{\bm{x}}^s} = {\bm{\bar x}} + {{{\bm{\hat x}}}^s}
    \label{eq:formulation}
\end{equation}
where ${{\bm{x}}^s}$ is the probabilistic forecasts, ${\bm{\bar x}}$ denotes the deterministic prediction results, expressed as ${\bm{\bar x}} = {f_\theta }_d\left( {{{\bm{x}}_\text{in}}} \right)$, where ${f_\theta }_d\left( {{{\bm{x}}_\text{in}}} \right)$ is a neural network with ${{{\bm{x}}_\text{in}}}$ as input and $\theta_d$ as its parameters. The generated prediction errors, ${{{\bm{\hat x}}}^s}$, follow the distribution ${{{\bm{\hat x}}}^s} \sim {G_\theta }_s\left( {{{{\bm{\hat x}}}^s}|\bm{y}} \right)$, where ${G_\theta }_s\left( {{{{\bm{\hat x}}}^s}|\bm{y}} \right)$ is a generative model conditioned on ${\bm{y}}$ with $\theta_s$ as its parameters.

\subsection{Score-based Diffusion Models for time series generation}
Score-based diffusion models \cite{song2019generative} have demonstrated significant success in the field of image generation. This paper investigates the application of score-based diffusion models for time series generation. As a generative model, the objective is to learn the underlying distribution of the target time series, denoted as $\bm{x}$. Once the distribution $p(\bm{x})$ is estimated, samples can be generated from it.
However, obtaining $p(\bm{x})$ is a challenging task. Score-based diffusion models overcome this difficulty by approximating the gradient of the log-likelihood function using a neural network, referred to as the \textbf{score}. Samples are then generated iteratively via Langevin dynamics, which can be mathematically expressed as follows:
\begin{equation}
{{\bm{x}}_{i - 1}} = {{\bm{x}}_i} + {\varepsilon _i}\nabla \log {p_i}\left( {{{\bm{x}}_i}} \right) + \sqrt {2{\varepsilon _i}} {{\bm{z}}_i}, i = K,K - 1,...,1
\end{equation}
where ${{\bm{x}}_K}$ is randomly sampled from a prior distribution (e.g., Gaussian). $\varepsilon _i$ is the sample step size. ${{\bm{z}}_i} \sim {{\cal N}}\left( {\bm{z}}_i; {\bm{0}}, {\bf{I}} \right)$ represents an additional noise term that ensures the generated samples do not collapse onto a single mode, but instead remain distributed around it, promoting diversity. The score function, $\nabla \log p_i(\bm{x}_i)$, defines a vector field across the entire space, indicating the direction in which the data space should be navigated to maximize the log probability.

The main limitation of this approach arises from the ill-defined nature of the score function when $\bm{x}$ lies on a low-dimensional manifold within a high-dimensional space, as off-manifold points have zero probability density. An effective solution to this issue is the introduction of multiple levels of Gaussian noise into the data. At each step of the noise addition process, the perturbation to the data remains minimal, gradually transforming it into a fully Gaussian distribution. This process is commonly referred to as diffusion \cite{ho2020denoising}. The sequence of progressively perturbed data distributions is as follows:
\begin{equation}
{p_{{t}}}\left( {{{\bm{x}}_t}} \right) = \int {p({\bm{x}})} {\cal N}\left( {{{\bm{x}}_t};l\left( {\bm{x}} \right),\sigma _t^2{\bf{I}}} \right)d{\bm{x}}
\end{equation}
Where $l(\bm{x})$ is linear with respect to $\bm{x}$, and $\left\{ {{\sigma _t}} \right\}_{t = 1}^T$ denotes a positive sequence of noise levels. A neural network ${s_\theta }({{\bm{x}}_t},t)$ is then trained using score matching to simultaneously learn the score function across all noise levels by minimizing the weighted Fisher divergence.
\begin{equation}
\mathcal{L}_{\text{SM}}(\theta) = 
\mathbb{E}_{\substack{t \sim U(0,T) \\ \bm{x}_t \sim p_t(\bm{x}_t)}} \left[ 
\lambda(t) \left\| s_{\theta}(\bm{x}_t,t) - \nabla_{\bm{x}_t} \log p_t(\bm{x}_t) \right\|^2 
\right]
\label{eq:LSM}
\end{equation}
where ${\lambda \left( t \right)}$ is a positive weighting function.

\subsection{Forward SDE and Reverse SDE}
The diffusion model involves two key processes: the diffusion process and the denoising process. Both can be described by stochastic differential equations (SDEs) when the noise in the perturbation is defined as a continuous-time variable $t \in [0, T]$. Let $\bm{x}(0)$ and $\bm{x}(T)$ represent the target time series and random noise, respectively, where $\bm{x}(0) \sim p_0$ and $\bm{x}(T) \sim p_T$. Here, $p_0$ denotes the distribution of the target time series, while $p_T$ represents a multivariate Gaussian distribution. The forward diffusion process is modeled as the solution to an It\^o SDE:

\begin{equation}
d{\bm{x}} = {\bm{f}}({\bm{x}},t)dt + g(t)d{\bm{w}}
\label{eq:forward}
\end{equation}
where $\bm{w}$ is the standard Wiener process, $\bm{f}({\bm{x}}, t)$ is a vector-valued function called the drift coefficient, and $g(t)$ is a scalar function known as the diffusion coefficient. The forward diffusion process transforms the distribution of target time series $p_0$ to a diffused distribution $p_T$.

\begin{figure}[htbp]
\centerline{\includegraphics[width=\columnwidth]{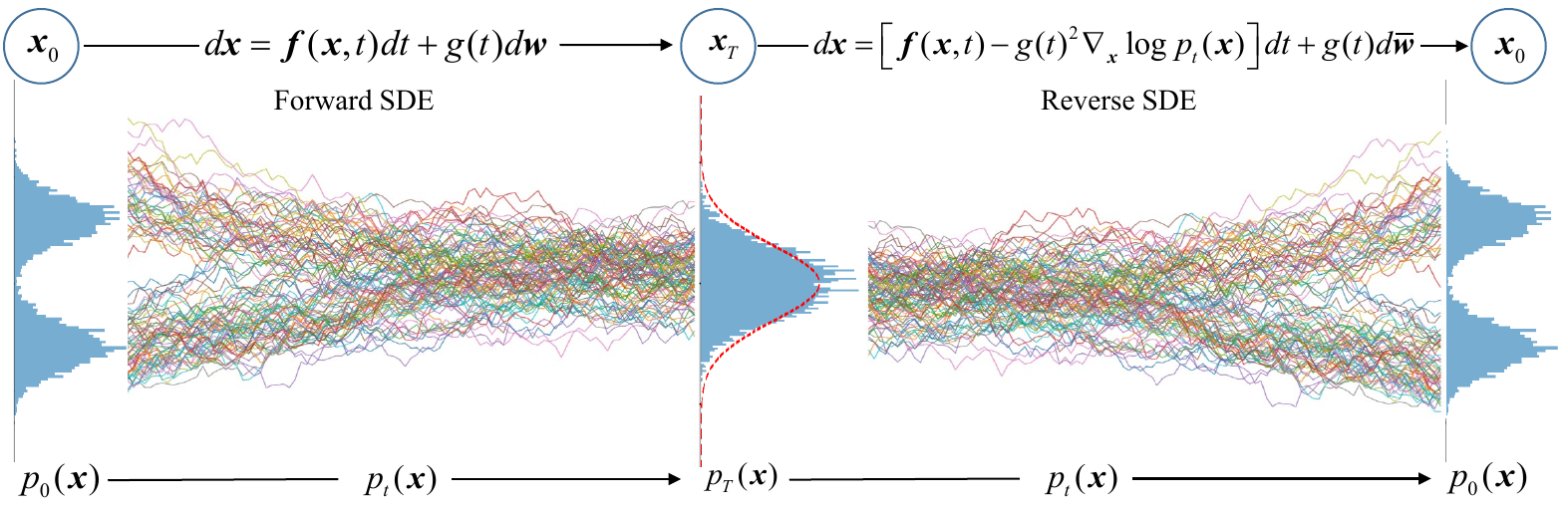}}
\caption{Overview of forward SDE and reverse SDE in score-based diffusion models.}
\label{fig:SDE}
\end{figure}

By starting from samples of \( \bm{x}(T) \sim p_T \) and reversing the diffusion process, one can obtain the target time series samples by sampling from \(p_0 \). The reverse of a diffusion process is also a diffusion process, running backwards in time and given by the reverse-time SDE:
\begin{equation}
    d{\bm{x}} = \left[ {f({\bm{x}},t) - g{{(t)}^2}{\nabla _{\bm{x}}}\log {p_t}({\bm{x}})} \right]dt + g(t)d{\bm{\bar w}}
    \label{eq:reverse}
\end{equation}
where ${{\bm{\bar w}}}$ is a standard Wiener process when time flows backwards from \( T \) to 0, and \( dt \) is an infinitesimal negative timestep. Once the score of each marginal distribution or conditional distribution, ${{\nabla _{\bm{x}}}\log {p_t}({\bm{x}})}$, is known for all \( t \), we can derive the reverse diffusion process from Eq.~\eqref{eq:reverse} and simulate it to sample from \( p_0 \). Fig.~\ref{fig:SDE} illustrates the framework of score-based diffusion models. One can map time series to a prior distribution with an SDE, and reverse it for generative models.

\section{Score-based Conditional Diffusion Models}
\label{sec:conditional}
\begin{figure}[htbp]
\centerline{\includegraphics[width=\columnwidth]{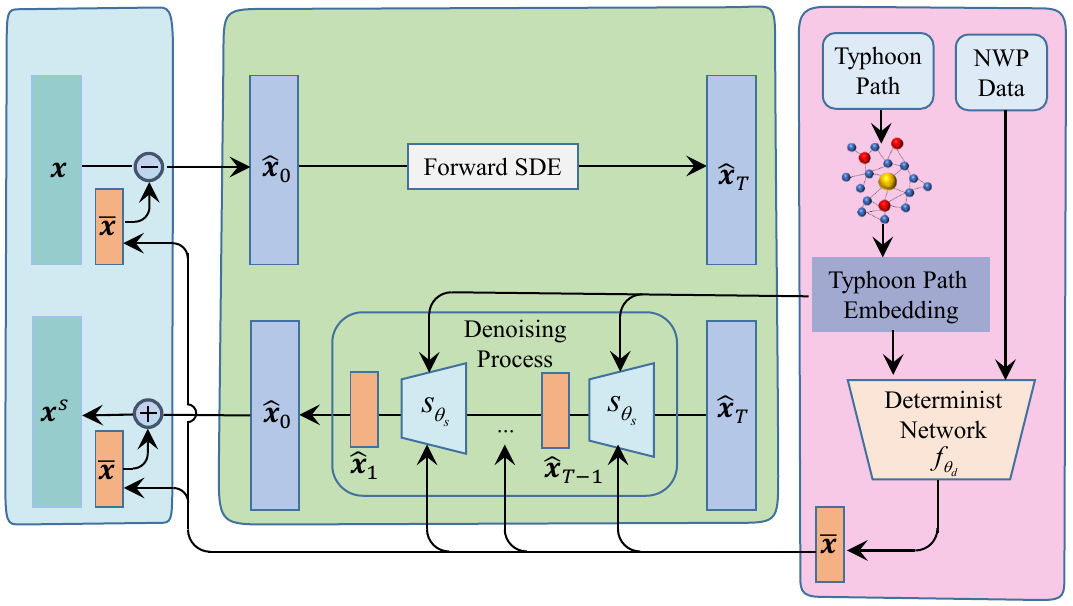}}
\caption{Framework of the SCDM for probabilistic forecasting under typhoon conditions.}
\label{fig:framework}
\end{figure}

The aim of this paper is to achieve WPF accurately under typhoon conditions. As shown in Eq.~\eqref{eq:formulation}, we approach this task by decomposing the predicted power into the sum of the deterministic prediction and the generated prediction error. The entire process is illustrated in Fig.~\ref{fig:framework}. First, the typhoon path forecast information is embedded in a knowledge graph to obtain the embedding vector of the typhoon path. This vector is then input into the deterministic network along with the NWP data to generate the deterministic wind power prediction $\bm{\bar{x}}$. Second, the typhoon path embedding vector and $\bm{\bar{x}}$ are used as known conditions, and a denoising network is employed to progressively map the random noise to the prediction error $\bm{\hat{x}}$ of wind power. Finally, the deterministic predictions are then combined with the sampled prediction errors to obtain probabilistic forecast $\bm{\hat{x}}^s$. Below, we will systematically walk through the relevant sections.

\subsection{Knowledge graph embedding of typhoon path}
\label{subsec:KG}
There is a scarcity of offshore wind power data affected by typhoons, and many typhoons have impacted the regions where wind farms are located prior to their establishment, a characteristic that can be utilized for data augmentation. In the context of knowledge graphs (KG), to address the issue of data sparsity, entities and relations are embedded into a low-dimensional continuous vector space using a technique known as knowledge graph embedding (KGE). Taking into account the error in typhoon path predictions, the coordinates of the typhoon center are discretized in $\ell$-degree latitude and longitude steps, and then mapped into a low-dimensional continuous vector space using the TransE algorithm \cite{bordes2013translating}. In TransE, a triple (\textit{head entity}, \textit{relation}, \textit{tail entity}) is generally denoted as $(\bf{e}_h, \bf{e}_r, \bf{e}_t)$, and if such a triple exists, the corresponding vectors in the vector space must satisfy the following relation:
\begin{equation}
    \bm{e}_h+\bm{e}_r \approx \bm{e}_t
    \label{eq:transE}
\end{equation}
The above equation means that when $(\bm{e}_h, \bm{e}_r, \bm{e}_t)$ holds, $\bm{e}_t$ should be a nearest neighbor of $\bm{e}_h+\bm{e}_r$, while $\bm{e}_h+\bm{e}_r$ should be far away from $\bm{e}_t$ otherwise. In this study, $\bm{e}_h$ denotes the embedding vector of coordinates of typhoon center. $\bm{e}_t$ denotes the embedding vector of offshore wind farms, and $\bm{e}_r$ is the embedding vector of relation between the typhoon and wind farm.

To learn such embeddings, we minimize the following loss function in the training data set:
\begin{equation}
   \mathcal{L}_{\text{KG}}(\theta_k) = \sum_{\substack{(\bm{e}_h,\bm{e}_r,\bm{e}_t) \\ (\bm{e}_h',\bm{e}_r,\bm{e}_t')}} [\gamma + \left\| {\bm{e}_h + \bm{e}_r - \bm{e}_t} \right\|_2^2 - \left\| {\bm{e}_h' + \bm{e}_r - \bm{e}_t'} \right\|_2^2]_+ 
\label{loss:transe}
\end{equation}
where $[x]_+$ denotes the positive part of $x$, $\gamma > 0$ is a margin hyperparameter. $(\bm{e}_h',\bm{e}_r,\bm{e}_t')$ is a random replacement for $(\bm{e}_h,\bm{e}_r,\bm{e}_t)$.
\subsection{Deterministic Network}
The proposed deterministic network, illustrated in Fig.~\ref{fig:determin}, is designed to capture the spatiotemporal correlations within offshore wind farms cluster. It integrates convolutional operations and attention mechanisms to jointly model local patterns and global dependencies. The architecture consists of the following components:  

1) Input layer: The model takes numerical weather prediction (NWP) data (e.g., pressure, wind speed) along with typhoon path embeddings, which are encoded from historical trajectory coordinates (e.g., latitude, longitude and intensity), as input. A linear transformation is applied to align the input dimensions.  

2) Feature extraction: The core architecture comprises four repetitive blocks. Each block employs a self-attention mechanism to capture complex spatiotemporal dependencies in offshore wind power. The 2D convolutional layers used to compute the query(Q), key(K), and value(V) matrices utilize non-uniform kernel sizes ($k=2,3,6,7$ in the 4 blocks, respectively), enabling hierarchical feature extraction across different spatial and temporal scales. Moreover, an adjacency matrix encoding the absolute distance between wind farms is directly incorporated into the attention score matrix to enhance spatial awareness. 
The Dis matrix is given by:
\begin{equation}
Dis_{i,j} = \left\{ {\begin{array}{*{20}{l}}
{\exp \left( { - {{\left( {{{\rm{dist}}(i,j)/\text{SD}}{ }} \right)}^2}} \right),}&{{\rm{if~dist}}(i,j) \ne 0}\\
{0,}&{{\rm{otherwise}}}
\end{array}} \right.
\end{equation}
where ${{\rm{dist}}}(i,j)$ is the distance between windfarm $i$ and $j$, $\text{SD}$ is standard deviation. 

3) Output layer: The feature representations from all blocks are concatenated and further refined through skip connections and layer normalization, which help mitigate gradient vanishing and accelerate convergence. Finally, a linear layer outputs the deterministic wind power predictions.  The loss function employs the mean square error (MSE) as the loss metric, which is defined as:
\begin{equation}
\mathcal{L}_{\text{MSE}}(\theta_d)=\mathbb{E}_{
        \bm{x} \sim p(\bm{x})
} \left[ 
    \left|| \bm{x} - \bm{\bar{x}} \right||_2^2
\right]
\label{loss:mse}
\end{equation}

While the deterministic network provides point estimates, they are inherently unable to quantify the uncertainty of the predictions. To address this limitation, uncertainty estimation is handled separately by the denoising network.

\begin{figure}[htbp]
\centerline{\includegraphics[width=\columnwidth]{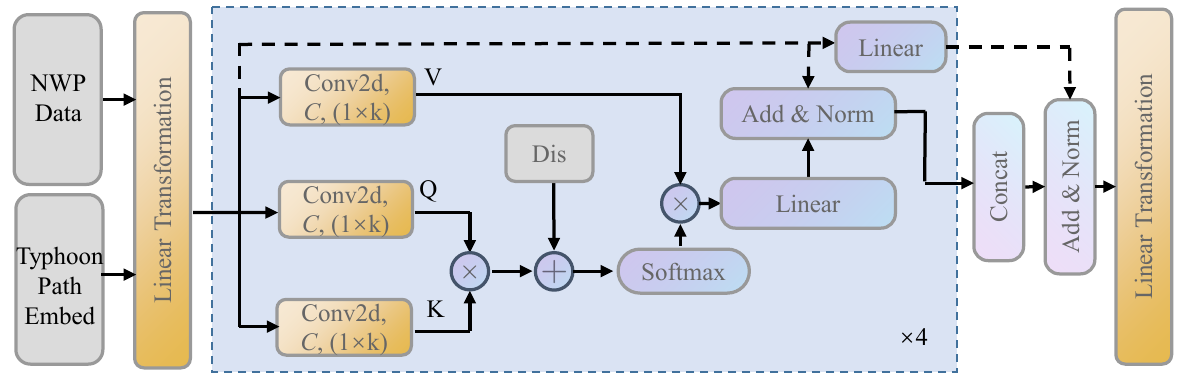}}
\caption{Structure of determinist network.}
\label{fig:determin}
\end{figure}

\subsection{Conditional denoising estimator (CDE) for generating prediction errors}
We use denoising networks to model the distribution of prediction errors. Eq.~\eqref{eq:LSM} presents the loss function for training an unconditional denoising network. However, direct optimization is not feasible, as the ground truth score \( \nabla_{\bm{x}_t} \log p(\bm{x}_t) \) is not accessible. Therefore, a loss function, as presented in Eq.~\eqref{eq:DSM}, is proposed in \cite{song2020score} and is mathematically equivalent to Eq.~\eqref{eq:LSM}. The ${{\nabla _{{{\bm{x}}_t}}}\log p(\bm{x}{_t}|{\bm{x}_0})}$ in Eq.~\eqref{eq:DSM} is analytically tractable, so the expectation can be approximated using Monte Carlo estimation.
\begin{equation}
\mathcal{L}_{\text{DSM}}(\theta_s) = \mathbb{E}_{
    \substack{
        t \sim U(0,T) \\
        \bm{x}_0 \sim p(\bm{x}_0) \\
        \bm{x}_t \sim p(\bm{x}_t | \bm{x}_0)
    }
} \left[ 
    \lambda(t) \left\| \nabla_{\bm{x}_t} \log p(\bm{x}_t | \bm{x}_0) - s_{\theta_s}(\bm{x}_t, t) \right\|_2^2 
\right]
\label{eq:DSM}
\end{equation}

Given the condition $\bm{y}$, and following a similar approach of using a neural network $s_\theta(\bm{\hat{x}}_t, \bm{y}, t)$ to approximate $\nabla_{\bm{\hat{x}}_t} \log p(\bm{\hat{x}}_t | \bm{y})$, the continuous denoising score-matching objective is as follows:  
\begin{equation}
\mathcal{L}_{\text{CDE}}(\theta_s) = \mathbb{E}_{\substack{
    t \sim U(0,T) \\
    \bm{\hat{x}}_t, \bm{y} \sim p(\bm{\hat{x}}_t, \bm{y})
}} \left[ 
    \lambda(t) \left\| \nabla_{\bm{\hat{x}}_t} \log p(\bm{\hat{x}}_t | \bm{y}) - s_{\theta_s}(\bm{\hat{x}}_t, \bm{y}, t) \right\|_2^2 
\right]
\label{eq:LCDE}
\end{equation}
The challenge associated with Eq.~\eqref{eq:LCDE} lies in the inability to compute the conditional scores explicitly. A feasible approach to address this issue is to expand the conditional scores using Bayesian formulas, as shown below.
\begin{equation}
\nabla_{\bm{\hat{x}}_t} \log p(\bm{\hat{x}}_t | \bm{y}) = \nabla_{\bm{\hat{x}}_t} \log p(\bm{\hat{x}}_t) + \nabla_{\bm{\hat{x}}_t} \log p(\bm{y} | \bm{\hat{x}}_t)
\end{equation}
However, this approach necessitates additional training of an auxiliary model to learn $\nabla_{\bm{\hat{x}}_t} \log p(\bm{y} | \bm{\hat{x}}_t)$. Therefore we exclude this approach from our study.

An alternative approach is to apply an equivalence transformation to Eq.~\eqref{eq:LCDE} in its entirety, which, through rigorous mathematical proof, leads to Theorem \ref{th:loss}, as presented in \cite{batzolis2021conditional}. From Theorem 1, it follows that the training process of the conditional score-based diffusion model closely resembles that of the unconditional score-based diffusion model, with the main difference being the inclusion of additional observed conditions as model inputs.
\begin{theorem}
The continuous denoising score-matching loss function $\mathcal{L}_{\text{CDE}}(\theta_s)$ in Eq.~\eqref{eq:LCDE} is also equal to:
\[
\mathbb{E}_{
    \substack{
        t \sim U(0,T) \\
        \bm{\hat{x}}_0,\bm{y} \sim p(\bm{\hat{x}}_0, \bm{y}) \\
        \bm{\hat{x}}_t \sim p(\bm{\hat{x}}_t | \bm{x}_0)
    }
} \left[ 
    \lambda(t) \left\| \nabla_{\bm{\hat{x}}_t} \log p(\bm{\hat{x}}_t | \bm{\hat{x}}_0) - s_{\theta_s}(\bm{\hat{x}}_t,\bm{y}, t) \right\|_2^2 
\right]
\]
\label{th:loss}
\end{theorem}

Based on Theorem \ref{th:loss}, the training approach of the unconditional score-based diffusion model can be directly replicated. The model can be trained using $\mathcal{L}_{\text{CDE}}(\theta)$ as the loss function for the denoising neural network $s_\theta(\bm{\hat{x}}_t,\bm{y}, t)$, without requiring additional auxiliary models. After training, the wind power prediction error under typhoon conditions can be sampled from $p(\bm{\hat{x}} | \bm{y})$ using the following reverse-time stochastic differential equation (SDE).
\begin{equation}
    d{\bm{\hat{x}}} = \left[ {f({\bm{\hat{x}}},t) - g{{(t)}^2}s_{\theta_s}(\bm{\hat{x}}_t,\bm{y}, t)} \right]dt + g(t)d{\bm{\bar w}}
\label{eq:conditional_reverse}
\end{equation}

\subsection{Mean Reverting SDE}
The score-based diffusion model offers flexibility in the choice of SDEs. In this work, we adopt a specific SDE that admits closed-form solutions, formulated as follows.
\begin{equation}
d{\bm{\hat{x}}} = {\alpha _t}\left( {{\bm{\mu }} - {\bm{\hat{x}}}} \right)dt + \sqrt {2{\alpha _t}} d{\bm{w}}
\label{eq:mean-reverse}
\end{equation}
The above SDE means that $f({\bm{\hat{x}}},t)$ and $g(t)$ have the following form.
\begin{equation}
\begin{array}{c}
f({\bm{\hat{x}}},t) = {\alpha _t}\left( {{\bm{\mu }} - {\bm{\hat{x}}}} \right)\\
g(t) = \sqrt {2{\alpha _t}} 
\end{array}
\end{equation}
Here, \( \bm{\mu} \) represents the state mean, and \( \alpha_t \) denotes a time-dependent positive parameter that governs the speed of mean reversion. In general, \( \bm{\mu} \) and the initial state \( \bm{\hat{x}}_0 \), given the condition \( \bm{y} \), can be any pair of distinct series. The forward SDE \eqref{eq:mean-reverse} then transforms one series into the other through a form of noise perturbation. For simplicity, we define \( \bm{\hat{x}}_0 \) as the ground truth wind power prediction error under typhoon conditions and set \( \bm{\mu} \) to zero.  

The solution to the mean-reverting SDE in \eqref{eq:mean-reverse} for \( \bm{\mu} = \bm{0} \) is given by  
\begin{equation}  
p(\bm{\hat{x}}_t \mid \bm{\hat{x}}_\tau) = \mathcal{N}\left( \bm{\hat{x}}_t ; \bm{\hat{x}}_\tau e^{-\bar{\alpha}_{\tau:t}}, \left(1 - e^{-2\bar{\alpha}_{\tau:t}}\right) \mathbf{I} \right)  
\label{eq:solution}  
\end{equation}  
where \( \bar{\alpha}_{\tau:t} := \int_\tau ^t \alpha_z dz \) are known coefficients. Further details on the derivation of this solution can be found in Appendix \ref{sec:solution}. Intuitively, from Eq.~\eqref{eq:solution}, when \( \tau = 0 \), as \( t \to \infty \), the term \( e^{- \bar{\alpha}_{\tau:t}} \) approaches zero, causing the conditional distribution \( p(\bm{\hat{x}}_t \mid \bm{\hat{x}}_\tau) \) to converge to a standard normal distribution ${{\cal N}}\left( {{\bf{0}},{\bf{I}}} \right)$. 

\subsection{Noise Matching Loss Function}
Based on Eq.~\eqref{eq:solution}, the conditional score can be expressed as:
\begin{equation}  
\nabla_{\bm{\hat{x}}_t} \log p(\bm{\hat{x}}_t \mid \mathbf{\hat{x}}_0) = -\frac{\bm{\hat{x}}_t - \bm{\hat{x}}_0 e^{-\bar{\alpha}_t}}{1 - e^{-2\bar{\alpha}_t}}
\label{eq:score}  
\end{equation}  
Meanwhile, using the reparameterization trick, Eq.~\eqref{eq:solution} can be reformulated as: 
\begin{equation}  
  \bm{\hat{x}}_t = \bm{\hat{x}}_0 e^{-\bar{\alpha}_t} + \sqrt{1 - e^{-2\bar{\alpha}_t}} \cdot \bm{z}_t
\label{eq:representation}  
\end{equation}  
where ${{\bm{z}}_t} \sim {{\cal N}}\left( {{{\bm{z}}_t};{\bf{0}},{\bf{I}}} \right)$. By combining Eq.~\eqref{eq:representation} and defining $\sigma _t^2:=1 - e^{-2\bar{\alpha}_t}$, the conditional score function can be rewritten as:
\begin{equation}  
\nabla_{\bm{\hat{x}}_t} \log p(\bm{\hat{x}}_t \mid \bm{\hat{x}}_0) = -\frac{\bm{z}_t}{\sigma_t}
\label{eq:score_z}  
\end{equation}  
To estimate the noise, we introduce a time-dependent neural network $s_\theta(\bm{\hat{x}}_t, \bm{y}, t)$ and optimize it with the noise matching loss:
\begin{equation}
\mathcal{L}_{\text{CDE}}(\theta_s)=\mathbb{E}_{
    \substack{
        t \sim U(0,T) \\
        \bm{\hat{x}}_0,\bm{y} \sim p(\bm{\hat{x}}_0,\bm{y}) \\
        \bm{\hat{x}}_t \sim p(\bm{\hat{x}}_t | \bm{\hat{x}}_0)
    }
} \left[ 
    \lambda(t) \left\| \frac{\bm{z}_t}{\sigma_t} + s_\theta(\bm{\hat{x}}_t,\bm{y}, t) \right\|_2^2 
\right]
\label{eq:LCDE_specify}
\end{equation}
In practice, we set the weighting function ${\lambda \left( t \right)}$ as $\sigma_t^2$, yielding the following practical form of Eq.~\eqref{eq:LCDE_specify}:
\begin{equation}
\mathcal{L}_{\text{CDE}}(\theta_s)=\mathbb{E}_{
    \substack{
        t \sim U(0,T) \\
        \bm{\hat{x}}_0,\bm{y} \sim p(\bm{\hat{x}}_0, \bm{y}) \\
        \bm{\hat{x}}_t \sim p(\bm{\hat{x}}_t | \bm{\hat{x}}_0)
    }
} \left[ 
    \left\| \bm{z}_t + {\sigma_t}s_\theta(\bm{\hat{x}}_t,\bm{y}, t) \right\|_2^2 
\right]
\label{eq:LCDE_pratical}
\end{equation}

\subsection{Denoising Network}
\begin{figure}[htbp]
\centerline{\includegraphics[width=\columnwidth]{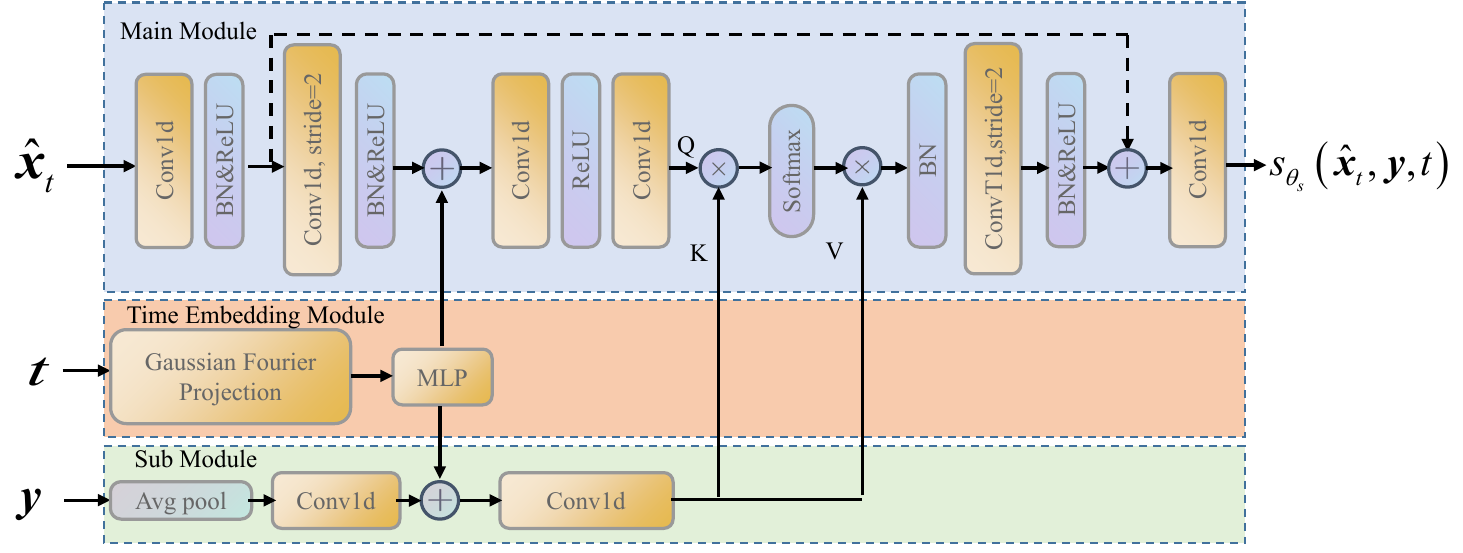}}
\caption{Structure of denoising network.}
\label{fig:denoising}
\end{figure}

Based on the loss function provided in Eq.~\eqref{eq:LCDE_pratical}, we have designed a network architecture that is naturally suited for generating wind power prediction errors under typhoon conditions. As illustrated in Fig.~\ref{fig:denoising}, the denoising network consists of three modules: the main module, the time embedding module, and the submodule. 

The main module is responsible for mapping the prediction error, perturbed by a multiscale perturbation kernel, into Gaussian random noise. Thus, this entire network can also be referred to as a noise prediction network. The main module preserves spatio-temporal consistency through mirrored downsampling/upsampling factors, while skip connections are employed to retain multi-resolution signal features.

The submodule primarily handles the typhoon condition information. Its input, \( \bm{y} = \text{concat}[\bm{\bar{x}}; \bm{e_t - e_h}] \), indicates that it is derived from a combination of the deterministic prediction results and the knowledge graph embedding of the typhoon path. The main module and the submodule perform feature fusion using a cross-attention mechanism. The submodule acts as a control valve, determining how much of the typhoon condition information should be incorporated into the prediction error in the main module.

The time embedding module is responsible for embedding the diffusion time step into both the main and submodules. The Gaussian embedding projection is given by \( \left[ \cos(2\pi wt); \sin(2\pi wt) \right] \), where \( w \) is a fixed parameter. The MLP is a three-layer neural network with Swish activation, designed primarily to adjust the dimensionality of the previous time projection.

\subsection{Algorithm of training and sampling}
After constructing the network architecture, training and inference (sampling) can be carried out using the corresponding data. The pseudo-code for the training process is presented in Algorithm \ref{alg:csdm}, which is divided into three main stages: the training of the TransE algorithm, the training of the deterministic network, and the training of the denoising network.

The pseudo-code for the probabilistic WPF process under typhoon conditions is presented in Algorithm \ref{alg:csdm_gen}. Consistent with the training process, the inference (sampling) process is also divided into three stages. The denoising process, modeled in continuous time, requires numerical discretization during sampling, with a discrete time step of $\Delta t$.

\begin{algorithm}[htbp]
\caption{Training procedure of the conditional score-based diffusion model}
\label{alg:csdm}
\KwIn{%
    The NWP data $\bm{x}_{nwp}$; the typhoon path data $\bm{x}_{typhoon}$; the measured wind power $\bm{x}$; margin $\gamma$; the speed of mean reversion $\alpha_t$; the learning rate \( \{lr_1, lr_2, lr_3\} \);.
}
\tcp{Typhoon path embedding training}
\While{$\mathcal{L}_{\text{KG}}(\theta_k)$ not converged}{

    \tcp{Forward propagation}
    sample a minibatch of $(\bm{e}_h,\bm{e}_r,\bm{e}_t)$
    
    construct negative triples $(\bm{e}_h',\bm{e}_r,\bm{e}_t')$
     
    calculate $\mathcal{L}_{\text{KG}}(\theta_k)$ with Eq.~\eqref{loss:transe} 
    
    \tcp{Back propagation}
    
    \( \theta_k \gets \text{Adam}(\nabla \mathcal{L}_{\text{KG}}(\theta_k), lr_1) \)
    
}
\tcp{Deterministic network training}
\While{$\mathcal{L}_{\text{MSE}}(\theta_d)$ not converged}{

    \tcp{Forward propagation}
    
    \( \bm{\bar{x}} = f_{\theta_d}(\bm{x}_{nwp},\bm{x}_{typhoon}) \) 
    
    calculate $\mathcal{L}_{\text{MSE}}(\theta_d)$ with Eq.~\eqref{loss:mse} 
    
    \tcp{Back propagation}
    
    \( \theta_d \gets \text{Adam}(\nabla \mathcal{L}_{\text{MSE}}(\theta_d), lr_2) \)
    
}

\tcp{Denoising network training}

\While{\(\mathcal{L}_{\text{CDE}}(\theta_s)\) not converged}{

    \tcp{Forward propagation} 

    calculate forecast errors $\bm{\hat{x}}_0 = \bm{x} - \bm{\bar{x}}$ 

    concatenate typhoon condition $\bm{y} = \text{concat}(\bm{e}_t-\bm{e}_h,\bm{\bar{x}})$
    
    sample \( t \sim \text{uniform}(0,T) \), calculate  \( \sigma _t^2=1 - e^{-2\bar{\alpha}_t} \)
    
    sample $\bm{z}_t \sim {{\cal N}}\left( {{\bf{0}},{\bf{I}}} \right)$

    calculate noisy targets $\bm{\hat{x}}_t$ with Eq.~\eqref{eq:representation}

    calculate $\mathcal{L}_{\text{CDE}}(\theta_s)$ with Eq.~\eqref{eq:LCDE_pratical}
    
    \tcp{Back propagation}
    
    \( \theta_s \gets \text{Adam}(\mathcal{L}_{\text{CDE}}(\theta_s), lr_3) \)

}

\Return{optimized parameters \( \theta_k^{*}, \theta_d^{*}, \theta_s^{*}\)}
\end{algorithm}

\begin{algorithm}[htbp]
\caption{Probabilistic forecasts process of the CSDM}
\label{alg:csdm_gen}
\KwIn{%
    The NWP data $\bm{x}_{nwp}$; the typhoon path data $\bm{x}_{typhoon}$; the number of samples \( S \); the speed of mean reversion $\alpha_t$; the optimized parameters \( \theta^*_k, \theta^*_d, \theta^*_s \).
}
\tcp{Forward TransE}
Indexing typhoon path vector embedding table, obtain $\bm{e_h}$ and $\bm{e_t}$

\tcp{Forward deterministic network}

The deterministic forecast: \( \bar{\bm{x}} \gets f_{\theta^*_d}([{\bm{x}_{nwp}};\bm{x}_{typhoon}]) \)

\tcp{Reverse SDE}

$\bm{y} = \text{concat}(\bm{e}_t-\bm{e}_h,\bm{\bar{x}})$

\For{\( s = 1, \ldots, S \)}{
    \( \hat{\bm{x}}_T \gets {{\cal N}}\left( {\bm{0}}, {\bf{I}} \right) \) \\
    \For{\( t = T, \ldots, 0 \)}{
    ${{{\bm{\hat x}}}_{t - \Delta t}} \gets {{{\bm{\hat x}}}_t} + \left[ {{\alpha _t}{{{\bm{\hat x}}}_t} + 2{\alpha _t}{s_{{\theta^{*} _s}}}\left( {{{{\bm{\hat x}}}_t},{\bm{y}},t} \right)} \right]\Delta t$
    
    sample $\bm{z} \sim {{\cal N}}\left( {\bm{0}}, {\bf{I}} \right)$

    ${{{\bm{\hat x}}}_{t - \Delta t}} \gets {{{\bm{\hat x}}}_{t - \Delta t}} + \sqrt {2{\alpha _t}\Delta t}  \cdot \bm{z}$
    
    }
    sample generation: \( \bm{x}^s \gets \hat{\bm{x}}_0 + \bar{\bm{x}} \)
}

\Return{sample set \( S = \{\bm{x}^1, \bm{x}^s, \ldots, \bm{x}^S\} \)}
\end{algorithm}

\section{Case Studies}
\label{sec:case_study}
\subsection{Data Collection and Preprocessing}
\label{sebsec:data}
The dataset used in this study is composed of three primary sources: the international best track archive for climate stewardship (IBTrACS) \cite{ibtracs2024}, historical wind power data from a cluster of offshore wind farms located along the coastal region of a province in southern China, and NWP data for each wind farm for the corresponding time periods. The IBTrACS dataset is primarily used to construct the knowledge graph of typhoon paths, as detailed in Section \ref{subsec:KG}. The other two datasets are mainly utilized for training and testing the deterministic and denoising networks.

The IBTrACS dataset is a global collection of tropical cyclone data available. For this study, we select typhoon paths of 39 typhoons that occurred between 2014 and 2024 to construct the knowledge graph, which assists in probabilistic WPF under typhoon conditions. For each typhoon, we extract the longitude, latitude, and intensity as the key features. The definition of the triples used in the knowledge graph is provided in Appendix~\ref{subsec:triples}.

\begin{figure}[htbp]
\centerline{\includegraphics[width=\columnwidth]{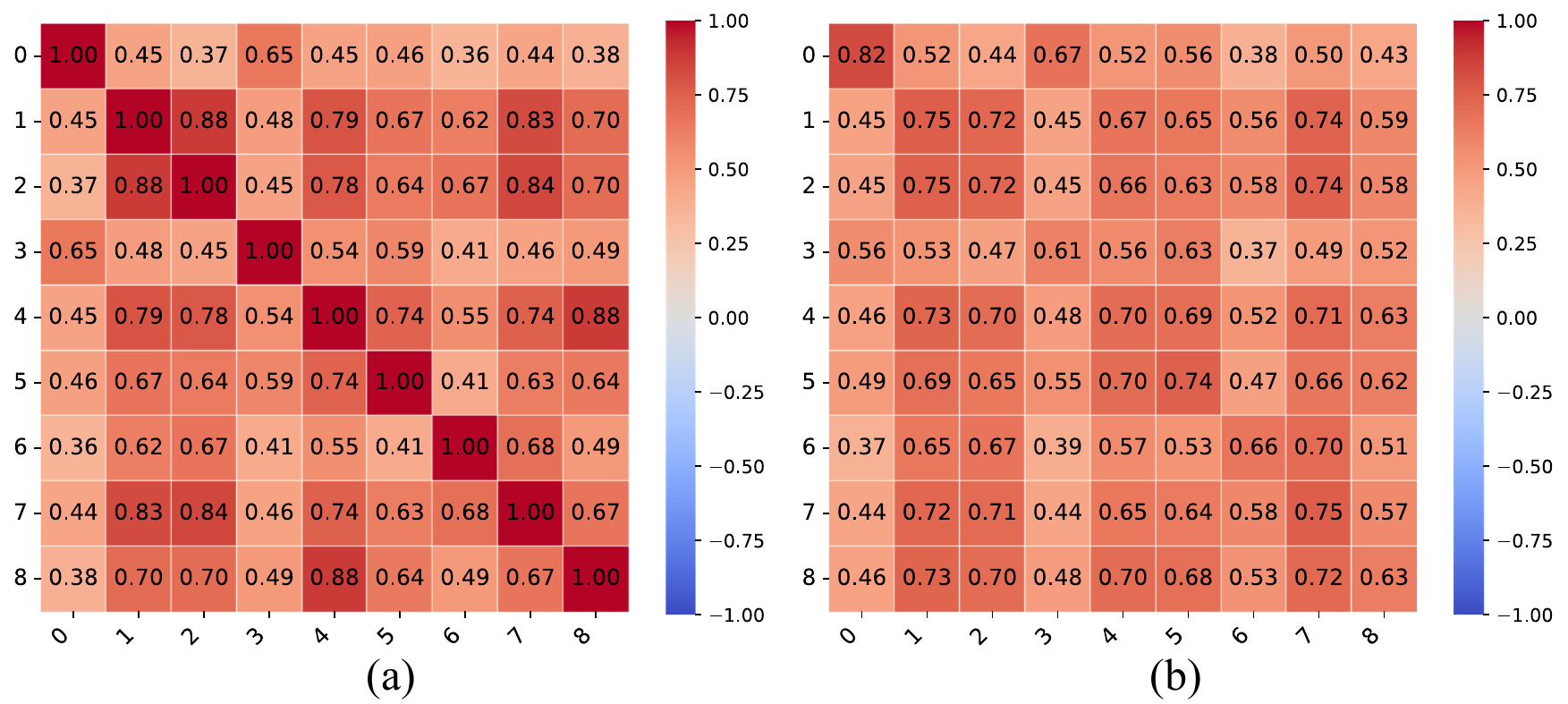}}
\caption{Correlation between forecasted wind speed and measured wind power among 9 offshore wind farms(a), Correlation of measured wind power between 9 offshore wind farms(b).}
\label{fig:corr}
\end{figure}

The historical wind power data span from August 6, 2022, to September 30, 2024, with a temporal resolution of 15 minutes. The wind farm cluster consists of 9 offshore wind farms, with a total installed capacity of 2,995 MW. Among these, the two wind farms farthest apart are separated by 800 km, with the power correlation between the farms depicted in Fig.\ref{fig:corr}(a). The smallest observed correlation coefficient is 0.36. In contrast, some wind farms are much closer to one another. For instance, \#1 and \#2 wind farms, as well as \#4 and \#8 wind farms, show a high wind power correlation of 0.88, as shown in Fig.\ref{fig:corr}(a). To normalize the data, the power output of each wind farm was divided by its installed capacity.

The NWP data includes forecast values for various weather elements at different altitudes. Since the NWP data at different altitudes for the same variable are highly linearly correlated, we selected only the wind speed, humidity, barometric pressure, and temperature data at a height of 100 meters. As shown in Fig.~\ref{fig:corr}(b), the correlation between the predicted wind speed and the measured wind power for each wind farm is relatively low, primarily due to the accuracy limitations of the NWP. To standardize the data, the NWP values were normalized by subtracting the mean and dividing by the standard deviation.

The computational experiments are conducted using the Pytorch framework on a two-GPU system with NVIDIA RTX 4090. For model optimization, we employ the Adam algorithm to implement stochastic gradient descent with an initial learning rate of $5\times10^{-4}$. During training, the learning rate is initially set at 0.0005 and gradually annealed to 0 following a cosine schedule.

\subsection{Baseline Settings}
To validate the effectiveness of our proposed method, we compare it with three deterministic prediction models and three probabilistic prediction models in this paper. For deterministic prediction, we select Informer \cite{zhou2021informer}, which employs ProbSparse self-attention and distilling operations to improve efficiency in long-sequence time-series forecasting; TimeXer \cite{wang2024timexer}, a state-of-the-art model ingesting external information to enhance the forecasting of endogenous variables; and Autoformer \cite{wu2021autoformer}, which introduces a decomposition architecture with auto-correlation mechanisms to enhance periodicity modeling. 

For probabilistic prediction, we include DeepAR \cite{salinas2020deepar}, an autoregressive RNN-based model generating probabilistic forecasts via likelihood estimation; CVAE \cite{sohn2015learning}, a conditional extension of variational autoencoders that integrates contextual information (e.g., historical patterns) through latent variables to guide uncertainty modeling; and WGAN \cite{gulrajani2017improved}, which leverages Wasserstein distance and adversarial training to stabilize the learning of data distributions. Experimental results demonstrate that our method achieves higher accuracy and greater stability across both deterministic and probabilistic evaluation metrics. It consistently provides reliable predictions, even in challenging scenarios with complex temporal patterns and limited observational data, such as the probabilistic forecasting of wind power under typhoon conditions.

\subsection{Performance Evaluation Metrics}

1) Deterministic Forecasting: The efficacy of deterministic forecasting is quantified through three principal evaluation metrics: mean absolute error (MAE), root mean square error (RMSE), and the coefficient of determination ($R^2$). These metrics are mathematically defined as follows.
\begin{equation}
\text{MAE} = \frac{1}{d} \sum_{i=1}^{d} \left| \bm{x}_i - {\bm{\bar {x}}}_i \right|
\label{eval:mae}
\end{equation}
\begin{equation}
\text{RMSE} = \sqrt{\frac{1}{d} \sum_{i=1}^{d} \left( \bm{x}_i - {\bm{\bar {x}}}_i \right)^2}
\label{eval:rmse}
\end{equation}
\begin{equation}
    {R^2} = 1 - \frac{{\sum\limits_{i=1}^{d} {{{\left( {\bm{x}_i - {\bm{\bar {x}}}_i} \right)}^2}} }}{{\sum\limits_{i=1}^{d} {{{\left( {\bm{x}_i - {\bm{x}}'_i} \right)}^2}} }}
    \label{R2}
\end{equation}
where $d$ is the dimension of $\bm{x}$ and $\bm{\bar {x}}$, ${\bm{x}}'$ is the mean value of $\bm{x}$. 

The MAE and RMSE respectively measure the average magnitude of absolute errors and squared deviations between predicted values and ground truth observations, with lower values indicating superior predictive accuracy. The $R^2$ statistic evaluates the proportion of variance in the observed data that is explainable by the forecasting model. Ranging from 0 to 1, the closer the value is to 1, the better the model’s predictions align with the actual observation.

2) Probabilistic forecasting: The Continuous Ranked Probability Score (CRPS), Energy Score (ES) and Variogram Score (VS) are used to evaluate probabilistic forecasts. These metrics assess both the accuracy and diversity of the generated samples.

If the predictions are \( S \) samples \( \{{\bm{x}}^{(i)} \mid i=1,\ldots,S\} \) from the forecast model, the CRPS can be approximately computed after sorting the samples:
\begin{equation}
\text{CRPS} \approx \frac{1}{S} \sum_{i=1}^S |x^{(i)} - x| - \frac{1}{2S^2} \sum_{i=1}^S \sum_{j=1}^S |x^{(i)} - x^{(j)}|
\label{eval:crps}
\end{equation}
where $x$ is the true observation. The CRPS quantifies the discrepancy between the predicted distribution and the true observed value, where smaller values indicate more accurate predictions.

Given a set of samples \( \{{\bm{x}}^{(i)} \mid i=1,\ldots,S\} \) and observations \( \bm{x} \), the ES is defined as:
\begin{equation}
\text{ES} = \frac{1}{S} \sum_{i=1}^{S} \left\| \bm{x} - {\bm{x}}^{(i)} \right\|_{2} 
          - \frac{1}{2S^{2}} \sum_{i=1}^{S} \sum_{j=1}^{S} \left\| {\bm{x}}^{(i)} - {\bm{x}}^{(j)} \right\|_{2}
\label{eval:es}
\end{equation}
where \( \left\| \cdot \right\|_{2} \) is the \( d \)-dimensional Euclidean norm. The ES compares the average distance between predicted samples and true observations, while penalizing internal dispersion between the predicted samples. Unlike CRPS, energy scores are designed for multivariate predictions and capture dependencies between variables. Lower ES values indicate better prediction performance.

The VS quantifies how well the differences between pairs of variables in the predictive distribution align with the corresponding differences in the true observations. It is defined as:
\begin{equation}
\text{VS} = \sum_{i,j=1}^{d} \left( |\bm{x}_{i} - \bm{x}_{j}|^{0.5} -  \frac{1}{S} \sum_{s=1}^{S} |\bm{x}_{i}^{(s)} - \bm{x}_{j}^{(s)}|^{0.5}\right)^2
\label{eval:vs}
\end{equation}
where $d$ is the dimension of $\bm{x}$ and $\bm{x}^{(s)}$. Lower values of VS indicate that the prediction more accurately reflects the structure between the variables.

\subsection{Typhoon Path Embedding}
We leverage knowledge graph embeddings of typhoon paths to facilitate probabilistic WPF for offshore wind farms under typhoon conditions. During the embedding process, only information related to typhoon paths—such as latitude, longitude, and intensity—is required. This enables the use of typhoon data from periods preceding the construction of offshore wind farms, thereby expanding the available dataset. As noted in Section \ref{sebsec:data}, typhoon records spanning more than a decade can be incorporated into the knowledge graph, regardless of the availability of corresponding wind power data during those earlier periods.

\begin{figure}[htbp]
\centerline{\includegraphics[width=0.7\columnwidth]{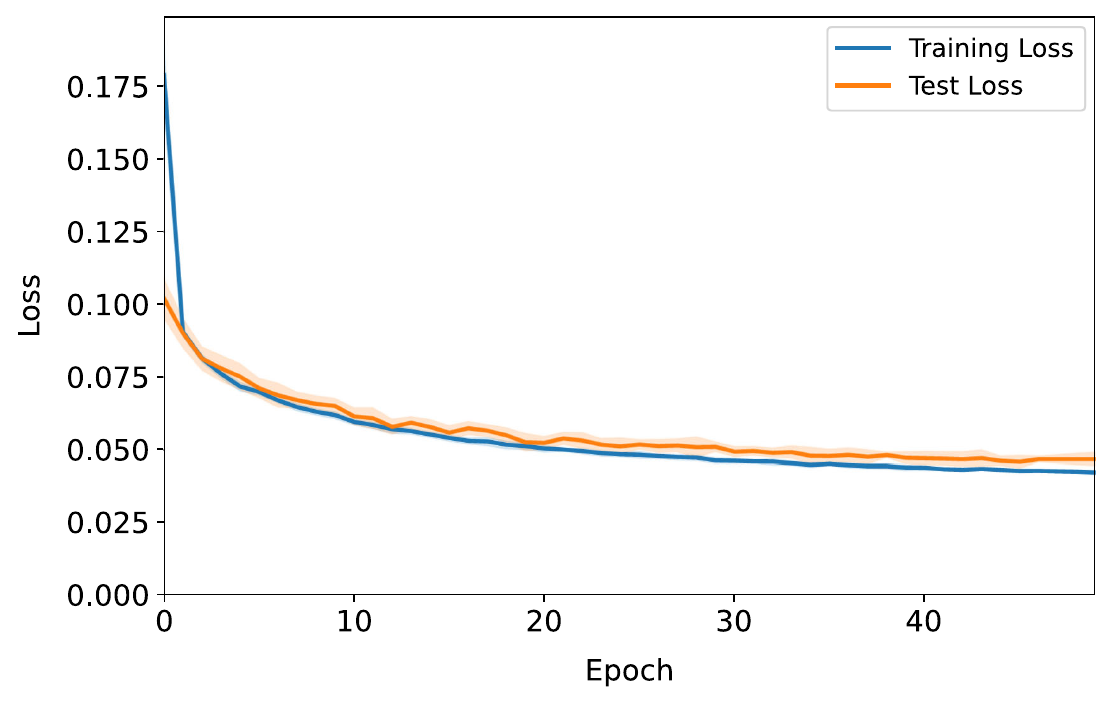}}
\caption{Loss value of the transE training process (dim=10).}
\label{fig:transE}
\end{figure}

To validate the effectiveness of TransE in capturing the semantic relationships between typhoon paths and wind farms, We visualize the training process and compare the deterministic prediction results with and without the incorporation of typhoon knowledge graph embeddings. As illustrated in Fig.~\ref{fig:transE}, the loss on both the training and validation sets progressively decreases and eventually stabilizes, indicating the successful convergence of the TransE algorithm.

To further demonstrate the practical utility of our approach, we select a test case from Typhoon "Maliksi" (2024). Fig.~\ref{fig:determin} compares the predicted and actual wind power curves for several offshore wind farms, with and without typhoon path embedding. In the figure, the blue line represents the true wind power values, the yellow line represents the predicted results from the deterministic network with typhoon path embedding, and the green line represents the predicted results from the deterministic network without typhoon path embedding.

The deterministic network with typhoon path embedding more accurately captures the rise and fall of wind power, as it takes into account the spatial semantics of the typhoon path encoded in the knowledge graph. As shown in Fig.~\ref{fig:determin}(d), the inclusion of typhoon path semantic information improves the accuracy of wind power prediction under typhoon conditions. Thus, typhoon path embedding proves valuable in enhancing our model's ability to capture temporal dependencies.

\begin{figure}[htbp]
\centerline{\includegraphics[width=\columnwidth]{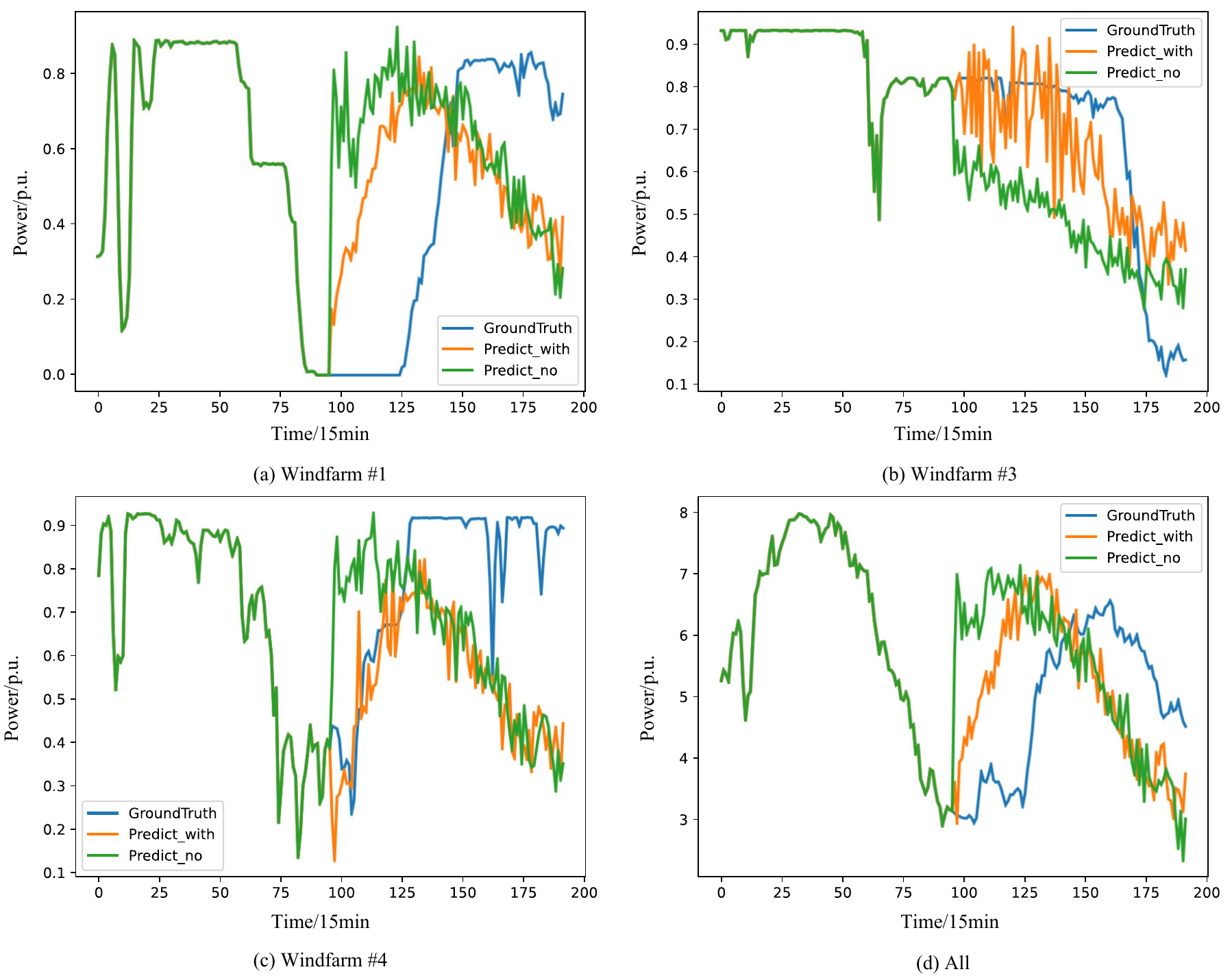}}
\caption{Comparison of deterministic prediction results with/without typhoon path embedding (dim=10).}
\label{fig:determin}
\end{figure}
\subsection{Performance comparison of deterministic forecasts}
In this section, we evaluate the prediction accuracy of the deterministic prediction models. We test the models on a dataset covering the period from May 10, 2024, to September 30, 2024, during which the offshore wind farm cluster was affected by two typhoons: Typhoon 'Maliksi' and Typhoon 'Yagi'. Table~\ref{tab:determ_normal} and Table~\ref{tab:determ_typhoon} present the average performance metrics over the entire test dataset, as well as specifically for the period during the two typhoons.

From table~\ref{tab:determ_normal}, the deterministic network ("Ours") achieves superior performance across all metrics for both 1-12h and 1-24h forecasts. Specifically, it reduces the MAE by 18.9\% (1-12h) and 11.9\% (1-24h) compared to the strongest baseline, Informer. From table~\ref{tab:determ_typhoon}, all models exhibit degraded performance due to extreme weather dynamics, yet our approach maintains its relative advantage. For 1-12h forecasts, our model achieves a 10.7\% reduction in MAE and a 13.8\% lower RMSE compared to the second-best model (TimeXer).

The performance advantage likely stems from our consideration of the typhoon path embedding and the attention mechanism, which can fully integrate information from exogenous factors—a critical capability for maintaining accuracy during both stable and extreme weather events.

\begin{table}[htbp]
\centering
\caption{Evaluation Metrics Results of Deterministic Forecasts under Normal Conditions}
\label{tab:determ_normal}
\resizebox{\columnwidth}{!}{%
\begin{tabular}{cccccccc}
\hline
\multirow{2}{*}{Model} & 1-12h &      &   &  & 1-24h &      &   \\ \cline{2-4} \cline{6-8} 
                       & MAE   & RMSE & $R^2$ &  & MAE   & RMSE & $R^2$ \\ \cline{2-8} 
Informer               & 0.0981     & 0.1401    & 0.6261 &  & 0.1161     & 0.1736    & 0.5638 \\
TimeXer                & 0.1099     & 0.1610    & 0.6041 &  & 0.1290     & 0.2005    & 0.5470 \\
Autoformer             & 0.1196     & 0.1784    & 0.5502 &  & 0.1313     & 0.2308    & 0.5214 \\
Ours                   & 0.0796     & 0.1380    & 0.6944 &  & 0.1023     & 0.1589    & 0.5712 \\ \hline
\end{tabular}%
}
\end{table}

\begin{table}[htbp]
\centering
\caption{Evaluation Metrics Results of Deterministic Forecasts under Typhoon Conditions}
\label{tab:determ_typhoon}
\resizebox{\columnwidth}{!}{%
\begin{tabular}{cccccccc}
\hline
\multirow{2}{*}{Model} & 1-12h &      &   &  & 1-24h &      &   \\ \cline{2-4} \cline{6-8} 
                       & MAE   & RMSE & $R^2$ &  & MAE   & RMSE & $R^2$ \\ \cline{2-8} 
Informer               & 0.2368     & 0.3721    & 0.3781 &  & 0.3279     & 0.4094    & 0.3613 \\
TimeXer                & 0.2189     & 0.3590    & 0.4058 &  & 0.3075     & 0.4003    & 0.3625 \\
Autoformer             & 0.2397     & 0.3902    & 0.3771 &  & 0.3495     & 0.4226    & 0.3415 \\
Ours                   & 0.1955     & 0.3093    & 0.4295 &  & 0.2983     & 0.3956    & 0.3691 \\ \hline
\end{tabular}%
}
\end{table}

\subsection{Performance comparison of probabilistic forecasts}
\begin{figure}
\centerline{\includegraphics[width=\columnwidth]{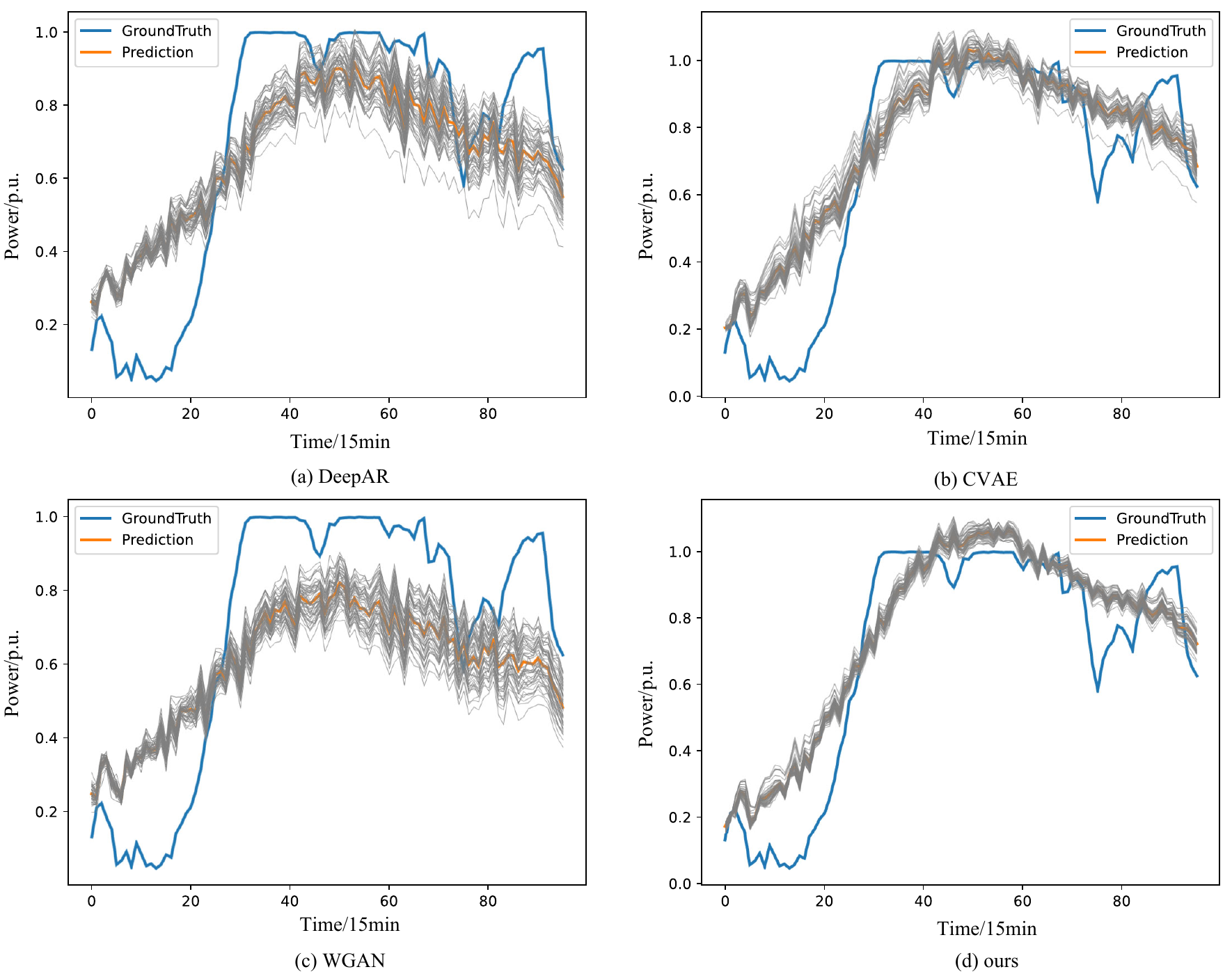}}
\caption{Comparison of probabilistic forecasts of windfarm \#0 during typhoon 'Yagi'(2024).}
\label{fig:probability}
\end{figure}

To generate more accurate prediction samples, we use the prediction results from the deterministic network as a baseline and then perform probabilistic generation of prediction errors. Fig~\ref{fig:probability} illustrates the prediction results of each probabilistic prediction model for offshore wind farm \#0 during Typhoon 'Yagi'. Each subfigure shows 50 generated samples (gray curves), with the orange curve representing the average of these samples and the blue curve indicating the true power. As shown in the figure, the samples generated by our proposed denoising network align more closely with the true power curve.
The performance metrics of samples generated for each probabilistic prediction model are summarized in Table~\ref{tab:Probability_typhoon}, where "ours" refers to our denoising network with the parameter set to $\alpha_t=0.1+19.9t,t\in [0,1]$. Our model achieves optimal performance across both 1-12h and 1-24h horizons, with the lowest scores in all metrics (1-12h: CRPS=0.0958, ES=9.1484; 1-24h: CRPS=0.1038, ES=14.0634). Our approach maintains consistent superiority in uncertainty quantification, particularly demonstrating enhanced robustness over extended horizons. The results highlight our architecture's effectiveness in modeling WPCSG. 
\begin{table}[htbp]
\centering
\caption{Evaluation Metrics Results of Probabilistic Forecasts under Typhoon Conditions}
\label{tab:Probability_typhoon}
\resizebox{\columnwidth}{!}{%
\begin{tabular}{cccccccc}
\hline
\multirow{2}{*}{Model} & 1-12h &      &   &  & 1-24h &      &   \\ \cline{2-4} \cline{6-8} 
                       & CRPS       & ES        & VS     &  & CRPS       & ES         & VS \\ \cline{2-8} 
DeepAR                 & 0.1256     & 11.4138    & 0.0214 &  & 0.1415     & 18.1679    & 0.0335 \\
CVAE                   & 0.1001     & 9.4760     & 0.0119 &  & 0.1070     & 14.4329    & 0.0256 \\
WGAN                   & 0.1377     & 12.0272    & 0.0242 &  & 0.1457     & 18.0217    & 0.0404 \\
Ours                   & 0.0958     & 9.1484     & 0.0105 &  & 0.1038     & 14.0634    & 0.0192 \\ \hline
\end{tabular}%
}
\end{table}

\section{Conclusion}
\label{sec:conclusion}
Probabilistic WPF under typhoon conditions presents significant challenges. In this study, we propose a novel score-based conditional diffusion model for probabilistic forecasting in offshore wind power cluster, specifically considering typhoon events. The proposed model demonstrates excellent deterministic and probabilistic prediction results by fully integrating external information. The effectiveness and superiority of the model are thoroughly analyzed and verified using real-world data from a provincial offshore wind farm cluster. Experimental results show that the proposed model consistently outperforms baseline models in terms of MAE, RMSE, CRPS, and ES.


\appendices

\section{Solution for the Mean-reverting SDE}
\label{sec:solution}

\textit{Proof for the solution of mean-reverting SDE.} 

Consider the mean-reverting SDE with \(\bm{\mu} = \bf{0}\):
\[
d\bm{x}_t = -\alpha_t \bm{x}_t dt + \sqrt{2\alpha_t} d\bm{w}_t
\]
where \(d\bm{w}_t\) is a Wiener process. Let \(\bar{\alpha}_{\tau:t} = \int_\tau^t \alpha_z dz\). We solve this linear SDE using the integrating factor method. Define the integrating factor:$F_t := e^{\bar{\alpha}_{\tau:t}}$, and apply Itô's lemma to \(d(\bm{x}_t F_t)\):
\[
d(\bm{x}_t F_t) = F_t d\bm{x}_t + \bm{x}_t dF_t + \underbrace{d[\bm{x}, F]_t}_{\bf{0}}
\]
Since \(dF_t = \theta_t F_t dt\) and \(F_t\) is deterministic, the cross-variation term vanishes. Substituting the SDE:
$$
\begin{aligned}
d(\bm{x}_t F_t) &= F_t \left(-\alpha_t \bm{x}_t dt + \sqrt{2\alpha_t} d\bm{w}_t\right) + \bm{x}_t \alpha_t F_t dt = F_t \sqrt{2\alpha_t} d\bm{w}_t
\end{aligned}
$$
Integrate from \(\tau\) to \(t\):
\[
\bm{x}_t F_t = \bm{x}_\tau F_\tau + \int_\tau^t F_s \sqrt{2\alpha_s} d\bm{w}_s
\]
Note \(F_\tau = 1\), we solve for \(\bm{x}_t\):
\[
\bm{x}_t = \bm{x}_\tau e^{-\bar{\alpha}_{\tau:t}} + \int_\tau^t e^{-\bar{\alpha}_{s:t}} \sqrt{2\alpha_s} d\bm{w}_s
\]
The mean and variance of $\bm{x}_t$ are as follows:
\[
\mathbb{E}[\bm{x}_t] = \bm{x}_\tau e^{-\bar{\alpha}_{\tau:t}} \quad \text{(Itô integral has zero expectation)}
\]
\[
\text{Var}(\bm{x}_t) = \mathbb{E}\left[\left(\int_\tau^t e^{-\bar{\alpha}_{s:t}} \sqrt{2\alpha_s} d\bm{w}_s\right)^2\right] = \left(1 - e^{-2\bar{\alpha}_{\tau:t}}\right) \mathbf{I}
\]
Thus, the transition probability is Gaussian:
\[
p(\bm{x}_t \mid \bm{x}_\tau) = \mathcal{N}\left( \bm{x}_t ; \bm{x}_\tau e^{-\bar{\alpha}_{\tau:t}}, \left(1 - e^{-2\bar{\alpha}_{\tau:t}}\right) \mathbf{I} \right)
\]

In summary, the entire solution process is completed.

\section{Definition of the Triple of Typhoon Paths}
\label{subsec:triples}
\textit{Head entity}: The coordinates of the typhoon paths are discretized into a 0.5-degree latitude/longitude grid. Considering that typhoons of varying intensities have different impacts on the same offshore wind farm, even when located at the same latitude and longitude, the head entity in the triplet is defined as the Cartesian product of the coordinates and the typhoon’s intensity.

\textit{Relationship}: The impact of a typhoon on a wind farm is influenced by both the distance to the typhoon center and its intensity. To model this, we discretize the distance between the typhoon center and the wind farm. When the distance is less than or equal to 350 km, the discretization step is set to 50 km. If the distance exceeds 350 km, it is assumed that the typhoon has no impact on the wind farm. As a result, there are 8 possible cases after discretizing the distance. The relationship in the triplet is defined as the Cartesian product of the discretized distance and the typhoon intensity.

\textit{Tail entity}: The wind farm identifiers are defined as the tail entity.


\bibliographystyle{IEEEtran}
\bibliography{reference.bib}

\end{document}